\documentclass[oldversion]{aa} 
\usepackage{graphicx}
\usepackage{natbib}
\bibpunct{(}{)}{;}{a}{}{,}

\begin{document}


\title{Radial velocity signatures of Zeeman broadening}

\author{A.~Reiners
  \inst{1}
  \and
  D.~Shulyak\inst{1}
  \and
  G.~Anglada-Escud\'e\inst{1}
  \and 
  S.V.~Jeffers\inst{1}
  \and
  J.~Morin\inst{1}
  \and 
  M.~Zechmeister\inst{1}
  \and\\
  O.~Kochukhov\inst{2}
  \and
  N.~Piskunov\inst{2}
}


\institute{Universit\"at G\"ottingen, Institut f\"ur Astrophysik, Friedrich-Hund-Platz 1, 37077 G\"ottingen, Germany\\
  \email{Ansgar.Reiners@phys.uni-goettingen.de} 
  \and 
  Department of Physics and Astronomy, Uppsala University, Box 516,
  751 20 Uppsala, Sweden }

\date{Received Sept 25, 2012 / Accepted Jan 14, 2013}


\abstract{ 

  Stellar activity signatures such as spots and plage can
  significantly limit the search for extrasolar planets. Current
  models of activity-induced radial velocity (RV) signals focused on
  the impact of temperature contrast in spots predicting the signal to
  diminish toward longer wavelengths. Contrary to this is the Zeeman
  effect on radial velocity measurements: the relative importance of
  the Zeeman effect on RV measurements should grow with wavelength
  because the Zeeman displacement itself grows with $\lambda$, and
  because a magnetic and cool spot contributes more to the total flux
  at longer wavelengths. In this paper, we model the impact of active
  regions on stellar RV measurements including both temperature
  contrast in spots and line broadening by the Zeeman effect. We
  calculate stellar line profiles using polarized radiative transfer
  models including atomic and molecular Zeeman splitting over large
  wavelength regions from 0.5 to 2.3\,$\mu$m. Our results show that
  the amplitude of the RV signal caused by the Zeeman effect alone can
  be comparable to that caused by temperature contrast; a spot
  magnetic field of $\sim$1000\,G can produce a similar RV amplitude
  as a spot temperature contrast of $\sim$1000\,K. Furthermore, the RV
  signal caused by cool \textit{and} magnetic spots increases with
  wavelength contrary to the expectation from temperature contrast
  alone. We also calculate the RV signal due to variations in average
  magnetic field strength from one observation to the next, for
  example due to a magnetic cycle, but find it unlikely that this can
  significantly influence the search for extrasolar planets. As an
  example, we derive the RV amplitude of the active M dwarf AD~Leo as
  a function of wavelength using data from the HARPS
  spectrograph. Across this limited wavelength range, the RV signal
  does not diminish at longer wavelengths but shows evidence for the
  opposite behavior consistent with a strong influence of the Zeeman
  effect. We conclude that the RV signal of active stars does not
  vanish at longer wavelength but sensitively depends on the
  combination of spot temperature and magnetic field; in active
  low-mass stars, it is even likely to grow with wavelength.

}


\maketitle
%

\section{Introduction}

The precise determination of radial velocities (RV) and their temporal
variations is a key data analysis method in stellar astrophysics. It
is applied to detect extra-solar planets and to measure their
projected mass, which requires high precision RV data pushing to the
order of m\,s$^{-1}$ and below \citep{2008PhST..130a4010M,
  2009IAUTA..27..316U}. Radial velocities can also be used to
determine periodic motion of the stellar surface enabling
asteroseismology to reveal sensitive information on fundamental
stellar parameters including a view into the interior of stars
\citep{2007A&ARv..14..217C, 2011A&A...526L...4B}.

Measurement of RV time series allows the detection of the projected
motion of a star or its surface from their spectral lines. For the
detection of planets, the assumption is that the shape of a spectral
lines does not vary with time so that its centroid position provides
information about the projected velocity of the star. The relative
radial velocity shift between two epochs is measured either by
searching for the best agreement between two spectra with radial
velocity as a free parameter, or by locating the centroid position in
a cross correlation profile calculated from the spectrum and some
template. Both methods assume that the spectral line shape is
identical. It is well known, however, that variations in the shape of
stellar line profiles on timescales similar to planetary orbits can be
caused by several mechanisms, e.g., by the transit of a planet
\citep{1924ApJ....60...15R, 1924ApJ....60...22M} or stellar
activity. The latter poses a number of problems to stellar radial
velocity measurements: first, active stars are typically fast rotators
implying wider line profiles reducing achievable accuracy in a given
radial velocity measurement
\citep[e.g.,][]{2001A&A...374..733B}. Second, activity is believed to
be caused by magnetic areas that can produce cool spots or hot plage
and are in general not symmetrically distributed over the star; this
allows the reconstruction of surface maps in tomographic imaging
studies \citep[e.g.,][]{1983PASP...95..565V}. Third, magnetic regions
can suppress convective motion and alter the signature of stellar
convective blueshifts \citep[see][]{2009ApJ...697.1032G,
  2010A&A...512A..39M}. Cool spots co-rotating with the stellar
surface are well studied for the Sun and much larger spots are known
to exist on other stars \citep{2005LRSP....2....8B,
  2009A&ARv..17..251S}. Line profile distortions due to these features
can lead to significant shifts of the line barycenter introducing an
offset to the real central line position
\cite[e.g.,][]{1998ApJ...498L.153S}. Useful information for the
correction of stellar RV curves for activity signatures can be
provided by activity indicators like the strength of chromospheric
emission lines or absorption line bisectors
\citep{2000A&A...361..265S}. Several examples exist where
periodicities in radial velocity curves were interpreted as due to
planetary companions but that later were retracted since the reason
for RV variability was found to be stellar activity. Additionally,
differences in RV amplitude between optical and infrared bands have
been detected in a few systems \citep[e.g.,][]{2006ApJ...644L..75M,
  2008A&A...489L...9H, 2011ApJ...736..123M}.

The effect of cool active regions on radial velocity measurements due
to temperature contrast alone has been investigated in detail by,
e.g., \citet{1997ApJ...485..319S, 2007A&A...473..983D,
  2009ApJ...707L..73M, 2010A&A...512A..38L, 2010ApJ...710..432R,
  2011MNRAS.412.1599B}. Magnetic flux tubes on the Sun are known to
exist also in the so-called network and plage areas
\citep[e.g.,][]{1984A&A...140..185S}. These regions show high magnetic
fields well above $B = 1$\,kG, temperature contrasts of a few hundred
K, and occupy a much larger fraction of the solar surface than cool
spots do. Due to their relatively low temperature contrast but large
filling factor, plage are difficult to characterize on surfaces of
stars other than the Sun. The impact of bright regions on RV
measurements (together with the impact of inhibited convection) was
investigated by \citet{2010A&A...512A..39M} using solar 2D images as
template for the spatial distribution of bright and dark areas. In
principle, the effect of bright areas on RV measurements is comparable
to the one of cool spots, but since flux contrast is reversed, the RV
signal has opposite sign. RV distortions from adjacent bright and dark
areas can therefore partially cancel out. The total contribution of
plage to the variability of solar irradiance is larger than the
contribution of cool spots \citep[e.g.,][]{1998GeoRL..25.4377F}, but
the two so far cannot be distinguished in any other star. Therefore,
plage and quiet stellar regions are often described by one single
atmospheric component, which probably is somewhat hotter than the
``quiet'' atmosphere. Line profile distortions in stars other than the
Sun are often described in a two-temperature model defining a cool
spot component in active stars. The real effect is likely a result of
spot and plage variability.

In general, co-rotating active regions can lead to spurious radial
velocity variations in phase with the stellar rotation period. The
amplitude of the variation depends on the projected rotational
velocity of the star, $v\,\sin{i}$, the fractional surface coverage of
the spotted area, $f$, the temperature contrast between ``quiet'' star
and active regions, and the wavelength used for observations. Since
the (absolute) flux ratio between active regions and photosphere is
smaller at longer wavelengths (independent of whether the active
region is cool or hot), it is expected that radial velocity
distortions due to stellar activity are lower at infrared than at
optical wavelengths. It has therefore been claimed that radial
velocity-based planet searches in very cool stars and brown dwarfs
(these objects are typically very active) should be optimally
conducted at infrared wavelengths rather than at optical
\citep[e.g.,][]{2006ApJ...644L..75M}.

The aim of this paper is to investigate the impact of magnetic fields
on the spectral appearance of active stars. It is not only the
temperature difference that affects stellar line profiles, but also
the field itself that may introduce significant distortions through
the Zeeman effect. The Zeeman effect operates opposite to temperature
contrast with more significant influence at longer wavelengths. Our
aim is to model the radial velocity signal of active regions including
both temperature and magnetic effects. After introducing our model
techniques and demonstrating the general picture in a few toy model
cases, we conclude that the Zeeman effects likely plays a very
significant role in the determination of radial velocities in active
stars.

\section{Zeeman splitting in stellar spectra}

Our picture of stellar surface properties is motivated by the
appearance of the Sun, where rising magnetic flux tubes inhibit
convection in active regions and produce both hot and cool surface
regions. Strongest flux concentrations are observed in cool spots with
typical temperatures several 100--1000\,K below photospheric average
and typical magnetic flux densities of several 100--1000\,G
\citep{2003A&ARv..11..153S}.

The general framework of stellar active regions is consistent with
observations from more active stars and stars of different spectral
type \citep{2009A&ARv..17..251S}. An important ingredient is that
active regions differ from the quiet photosphere in both temperature
and magnetic properties; in particular, large starspots are believed
to be both cool \emph{and} magnetic. Owing to our limited ability to
measure localized magnetic fields in other stars, however, we lack
good understanding of magnetic fields in stars other than the Sun. In
particular, we have no empirical evidence of the relation between
magnetic field and spot temperature in very active stars.
Nevertheless, we may find it reasonable to assume that local magnetic
field strengths in other stars are on the same order as in the Sun,
and that large active regions can be similar to very large
sunspots. Empirical results on average magnetic fields in very active
low-mass stars are consistent with this picture finding very strong
average surface fields at the kilo-Gauss level in mid- and late-M
dwarfs \cite[e.g.,][]{2007ApJ...656.1121R}.

The appearance of a spectral line in the presence of a magnetic field
is determined by the Zeeman effect: each energy level with a total
angular momentum quantum number $J$ splits into ($2J + 1$) states of
energy with different magnetic quantum numbers $M$. In absence of a
magnetic field, the transition energy is unique but it splits into
three groups of transitions according to the change in the magnetic
quantum number $M$ invoked by the transition ($\Delta M = -1, 0,
+1$). The appearance of the spectral line also depends on the geometry
of the field, but this effect is often neglected assuming an
``homogeneous'' distribution of field lines over the stellar
surface. We refer to \cite{2012LRSP....9....1R} for a deeper
discussion of magnetic field observations.

For the context of radial velocity measurements, we are interested in
the amplitude of spectral line deformations caused by magnetic
fields. The two $\sigma$-groups with magnetic quantum numbers $M = -1,
+1$ are shifted with respect to the $\pi$-group ($M = 0$) by an amount
that depends on the level's quantum numbers, condensed in the
Land\'e-factor $g$, and is proportional to the magnetic field $B$. In
velocity units, the displacement can be equated as
\begin{equation}
  \label{eq:deltav}
  \Delta v_{\rm Zeeman} = 1.4\, g\,\lambda\,B,
\end{equation}
with $v$ in m\,s$^{-1}$, $\lambda$ in $\mu$m, and $B$ in Gauss. The
Land\'e-factor $g$ is of order unity. Eq.\,\ref{eq:deltav} has two
important implications: 1) The typical displacement of Zeeman
components in the presence of magnetic fields is on the order of
1\,m\,s$^{-1}$\,G$^{-1}$. For typical field strengths of solar active
regions (100--1000\,G), flux from active regions can therefore be
displaced, through the Zeeman effect, by several hundred m\,s$^{-1}$;
2) in velocity units, the displacement is proportional to wavelength,
$\lambda$, of the spectral line under consideration. Thus, the
displacement is larger for longer wavelengths, which is contrary to
the displacement from temperature contrast \citep[see,
e.g.,][]{2010ApJ...710..432R}. We note that RV signatures due to
temperature contrast always diminish with wavelength independent of
whether they are hotter or cooler than the rest of the star. The
translation from the displacement of Zeeman components, $\Delta v_{\rm
  Zeeman}$, into a shift of the line profile barycenter, $\Delta
v_{\rm vrad}$, is non-trivial and subject of this paper. The typical
amplitude of $\Delta v_{\rm Zeeman}$ together with typical field
strengths on stars show that even if the net signal ($\Delta v_{\rm
  vrad}$) in a spectral line only would be a few percent of $\Delta
v_{\rm Zeeman}$, this still would easily be in the range relevant for
detecting planetary orbits through the RV method.

For the purpose of this paper, we are interested in the effect of
Zeeman splitting on stellar radial velocity curves. A constant average
magnetic field may affect the overall shape of a line profile with
respect to the non-magnetic case, but as long as this profile is not
time-variable, it is not relevant for radial velocity analysis. On the
other hand, any time-variability in the magnetic properties of stellar
surfaces can have significant consequences for stellar radial velocity
curves. Variability can be caused either by localized magnetism on
time-variable (projected) surface areas, like co-rotating spots, or by
intrinsic time-variability of the magnetic field observed on the
visible hemisphere (e.g., magnetic cycles). We will consider both
types of line profile variability in the following.

\section{Co-rotating magnetic spots}
\label{sect:Toy}

\begin{figure}
  \centering \mbox{
    \includegraphics[width=.45\textwidth]{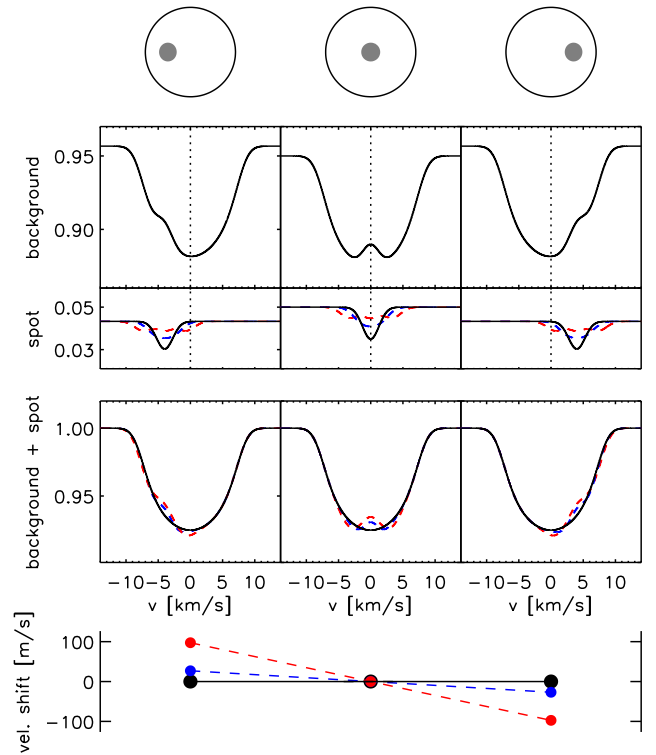}}
  \caption{\label{fig:Sketch}Sketch of the effect of a magnetic spot
    on radial velocity measurements. Top: rotating star with large
    spot, covering 5\% of the visible surface if observed at disk
    center, seen at three phases. The star is rotating with the spot
    approaching in the left column, centered in the middle column, and
    rotating out of view in the right column. Second row: stellar line
    profile broadened by rotation but without spot flux, showing the
    impact of a dark star spot. Third row: flux emerging from spot
    only, assuming same temperature as photosphere. Row four: sum of
    surrounding plus the spot region. Row five: Radial velocity
    determined from fitting a Gaussian profile to the spectral lines
    in row four. In rows three to five, black solid lines show the
    case for no magnetic field (and zero spot temperature
    contrast). Blue and red dashed lines show cases in which the spot
    area has $B = 1$\,kG and 2\,kG, respectively ($\lambda =
    1.2\,\mu$m and $g=1.0$). }
\end{figure}

Spectral lines of rotating stars are broader than lines of
non-rotating stars because of the Doppler effect: light emerging from
the area of the star rotating towards us is blue-shifted while light
coming from the area moving away from us is red-shifted. The net
effect is a characteristic spectral line broadening allowing precise
measurement of the star's projected rotational velocity, $v\,\sin{i}$
\cite[e.g.,][]{2005oasp.book.....G}. If a certain area of a star is
cooler, this area emits less flux than other regions. In absorption
lines, this leads to a characteristic bump at the position in the line
profile corresponding to the local velocity projected towards the
observer. In a similar fashion, the flux emitted from co-rotating
magnetic spots will alter the appearance of a Doppler-broadened line
profile, and the shape of the overall profile will change when the
star is observed at different rotational phases.

Stellar convective blueshift is another result of Doppler broadening
in spectral lines. Due to the imbalance between rising, hot plasma and
cooler downflows, spectral lines are generally blue-shifted in
sun-like stars \cite[e.g.,][]{2009ApJ...697.1032G}. In magnetic areas,
convective motion is suppressed, which can result in significant RV
signatures. \citet{2010A&A...512A..39M} have modeled this effect
assuming that in active regions the average convective blueshift seen
in all lines is attenuated by $\Delta v = 190$\,m\,s$^{-1}$
perpendicular to the solar surface. \citet{2010A&A...512A..39M} find
that in the Sun, the RV signature from convective blueshifts dominates
the activity-induced stellar RV signal; with an amplitude of several
m\,s$^{-1}$, the convective RV signal is larger than the signal due to
flux contrast by a factor of a few. In our study, we do not take into
account the signal from convective blueshifts because we are
concentrating on the additional effect of Zeeman splitting. Following
\citet{2010A&A...512A..39M}, one can argue that the RV variations due
to blueshift can be separated from the effects of the line shape (flux
contrast and Zeeman splitting).

\subsection{Toy model}

In the following, we calculate a line profile for an artificial star
rotating at $v\,\sin{i} = 2$\,km\,s$^{-1}$ and simulate a line profile
distortion due to an artificial spot with the same temperature as the
quiet photosphere. The distortion is only due to Zeeman splitting of
the line emerging from the spot. In our examples, we calculate the
rotational phases corresponding to maximum displacement of the line
center, i.e., we compute the semi-amplitude $K$ of the apparent radial
velocity curve due a magnetic spot \cite[see Fig.\,8
of][]{2010ApJ...710..432R}. The radial velocity is then calculated
from cross-correlating the undistorted template spectrum with the
spectrum of the spotted star. The barycenter of the cross-correlation
function is determined by fitting it with a Gaussian profile.

The top panel in Fig.\,\ref{fig:Sketch} shows a sketch of rotating
star at three rotational phases. The stellar surface shows a large
magnetic spot that is rotating into view in the left column, is
centered in the middle column, and is rotating out of view in the
right column. The spot covers 5\% of the visible surface at the center
of the disk but it appears smaller if viewed closer to the limb. The
second (from top) row of Fig.\,\ref{fig:Sketch} shows a stellar line
profile broadened by rotation with projected rotation velocity
$v\,\sin{i} = 2$\,km\,s$^{-1}$ but without any flux coming from the
spot region; the total flux is reduced by 5\% in this (central)
example. This shows the typical line profile when looking at the
temperature effect only. The third row shows the flux emerging from
the spotted area only, assuming it has the same temperature as the
surrounding photosphere. In general, large sun- or starspots are
believed to be cooler than the surrounding atmosphere, but we look at
the case of identical temperatures first to isolate the impact of the
Zeeman effect from the temperature contrast. Row four shows the sum of
the surrounding region (star without spot) plus the spot region. In
panels three and four, black solid lines show the case for no magnetic
field. Here, the profile is just the undistorted spectral line. Dashed
blue and red lines show cases in which the spot area harbors a
magnetic field with an average field strength of 1\,kG and 2\,kG,
respectively. A spectral line at $\lambda = 1.2\,\mu$m and $g=1.0$ is
assumed.  The spectral line emerging from the spot region is broader
and shallower in the magnetic case. In row four, the effect of one
co-rotating magnetic spot (again, without any temperature difference)
on a spectral line is displayed. Clearly, the Zeeman effect
significantly distorts the line profile and consequently shifts the
apparent center of the line. Finally, in the bottom panel, we quantify
this by showing the center of the spectral line as derived from a fit
assuming a Gauss function. We note that the radial velocity shift is a
consequence of the non-axisymmetric field distribution implying that
polar spots or other axisymmetric configurations cannot introduce RV
shifts through this mechanism (but see
Section\,\ref{sect:symmetric}). We can compare Fig.\,\ref{fig:Sketch}
of this work to Fig.\,8 in \citet{2010ApJ...710..432R} to see that the
line profile deformation induced by the Zeeman effect is similar to a
deformation induced by cool spots. In particular, the radial velocity
signal from a magnetic spot through the Zeeman effect has the same
sign as the signal from a cool spot due to flux contrast.

The amplitude of the apparent radial velocity shift in our example is
approximately 100\,m\,s$^{-1}$ for $B = 2$\,kG, and approximately
25\,m\,s$^{-1}$ for $B = 1$\,kG. It is worth noting that the projected
rotational velocity chosen for this example, $v\,\sin{i} =
2$\,km\,s$^{-1}$ is similar to solar rotation, which is considered
rather slow compared to typical late-type stars. Nevertheless, the
amplitude is significant for the precision required for the RV
accuracy level needed for planet search. For more rapidly rotating
stars the distortion scales with $v\,\sin{i}$
\citep[cp.][]{2010ApJ...710..432R}. Because of the similarities
between temperature and Zeman RV signatures, line profile diagnostics
like bisectors \citep{2007A&A...473..983D} can be useful tools to
investigate activity-related reasons for velocity shifts in observed
stellar spectra.

\subsection{Dependence on wavelength and field strength}

\begin{figure}
  \centering 
    \includegraphics[width=.475\textwidth]{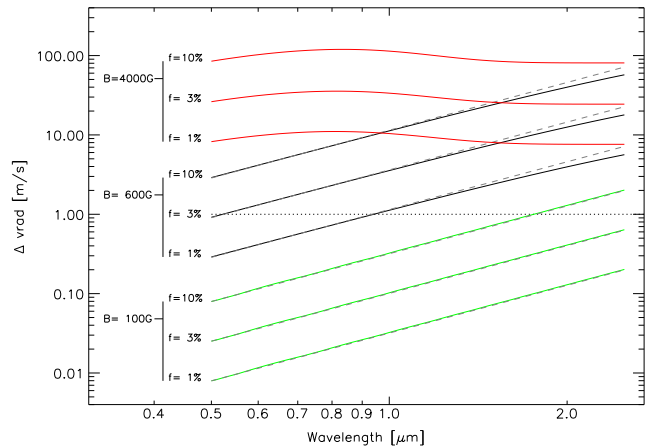}
    \caption{\label{fig:RVcorotation_Toy}Toy model radial velocity
      signal due to the Zeeman effect from a co-rotating magnetic spot
      with zero temperature contrast. The signal is calculated for
      three different field strengths inside the spot (green: 100\,G,
      black: 600\,G, and red: 4000\,G); for each case three different
      spot sizes are used ($f = 1\,\%, 3\,\%$, and $10\,\%$). Grey
      dashed lines show an analytical approximation of the scaling
      using Eq.\,\ref{eq:toyscaling}. The dotted line visualizes the
      1\,m\,s$^{-1}$ limit.}
\end{figure}

In Fig.\,\ref{fig:RVcorotation_Toy} we show the RV amplitude due to
the Zeeman effect calculated from one spectral line with $g = 1.0$ as
a function of wavelength. The line distortion is induced by a magnetic
spot with one out of three different values of magnetic field strength
inside the spot ($B = 100$\,G, 600\,G, and 4000\,G) and one out of
three different spot filling factors ($f = 1\,\%, 3\,\%$, and
$10\,\%$). While not much is known about the geometric concentration
of small magnetic areas on cool stars, the total magnetic energy
assumed in our examples is easily justified by observations of cool
star magnetic fields \citep{2012LRSP....9....1R}; our example stars
have average fields of $Bf = 1$--400\,G (concentrated in one single
spot) well in the range of average fields observed that can be as
strong as several kG \citep{2009ApJ...692..538R}.

In all cases, the apparent radial velocity shift scales with spot size
and higher field strength introduces larger radial velocity shift. For
relatively weak fields ($B \la 1$\,kG), the signal also grows with
wavelength as can be expected since the Zeeman effect does. The
amplitude of the RV signal during stellar rotation due to the
signature of the spot can be approximated by the following scaling
relation:
\begin{equation}
  \label{eq:toyscaling}
  \Delta v_{\rm{rad, toy}} = \mathrm{const} \times \, \,f \, (B \lambda)^{2},
\end{equation}
with $f$ the relative fraction of the spot area, $B$ the magnetic
field inside the spot, and $\lambda$ the wavelength. The grey dashed
lines in Fig.\,\ref{fig:RVcorotation_Toy} illustrate the scaling of
Eq.\,\ref{eq:toyscaling}; the radial velocity shift is proportional to
the filling factor, and the dependence on $B$ is identical to the one
on $\lambda$, which is consistent with Eq.\,\ref{eq:deltav}.

The amplitude of the radial velocity shift is below 1\,m\,s$^{-1}$ as
long as the field inside this one spot is on the order of $B =
100$\,G. In this case, only relatively high spot coverage (10\,\%) at
long wavelengths ($\lambda > 2\,\mu$m) can cause signals larger than
$\sim$1\,m\,s$^{-1}$. For the Sun we know that field strengths inside
a spot can easily be larger than 100\,G (but note that spots with
large fields are typically cool, which is not considered in this
simple model). In our simulation, spots with several hundred Gauss
field strengths and filling factors of a few percent can introduce
radial velocity signals well above the m\,s$^{-1}$ level. For example,
the signal of a spot with $f = 3\,\%$ and $B = 600$\,G causes a radial
velocity signal with an amplitude of $\Delta v_{\rm rad} \approx
1$\,m\,s$^{-1}$ at $\lambda = 500$\,nm. The signal grows with
wavelength up to $\Delta v_{\rm rad} \approx 10$\,m\,s$^{-1}$ at
$\lambda = 2200$\,nm.

The case of very strong magnetic fields (4000\,G) inside the spot
region shows a somewhat different behavior. While the radial velocity
shift in this case is larger than in the other cases with weaker
fields, it does not scale with wavelength and remains almost at a
constant value between 10 and 100\,m\,s$^{-1}$ depending on filling
factor. We interpret this behavior as a saturation effect in the sense
that the line profile distortion due to a spot with $B \approx
4000$\,G does not distort a measurement of the line center much more
than a spot with $B \approx 1000$\,G does. The reason for this is that
Zeeman broadening already is so significant that the essential effect
in the line profile is similar to a very cool spot in which the flux
from the spot area is simply missing. For a field strength of $B =
4000$\,G, the displacement of $\sigma$-components is $v_{\rm{Zeeman}}
\sim 6$\,km\,s$^{-1}$ (Eq.\,\ref{eq:deltav}), which means that flux
from the spot area is essentially removed to the wings of the spectral
line (note that the typical line width in Fig.\,\ref{fig:Sketch} is
$\sim 10$\,km\,s$^{-1}$). In other words, the radial velocity signal
does not grow any further as soon as the amplitude of Zeeman splitting
from the spot area is comparable to the line width of the rotating
star.

The conclusion from this exercise is that the Zeeman signal introduced
by a magnetic spot that is not cooler than the rest of the star can be
significant for radial velocity surveys aiming at precisions on the
order of a meter per second. The RV signature has the same sign as the
signature of a cool (non-magnetic) spot. The Zeeman signal grows with
wavelength for moderate values of $B$ inside the spot. The effect of a
(non-magnetic) \emph{cool} spot also can be very significant on the
m\,s$^{-1}$ level, but it scales with opposite sign, i.e., it is large
at short wavelengths but diminishes towards longer. For a similar
simulation using a toy model of (non-magnetic) cool spots, we refer to
\cite{2010ApJ...710..432R}. In their Fig.\,10, they show that in a
cool star ($T \sim 4000$--6000\,K), the effect of a $f = 2\,\%$ spot
that is $\Delta T = 200$\,K cooler than the surrounding is on the
order of 10\,m\,s$^{-1}$ at $\lambda = 550$\,nm, and the amplitude is
a factor of 3 smaller at $\lambda = 1800$\,nm. Furthermore, they show
that the RV amplitude is larger if temperature contrast is larger
(cooler spots in that example), but dependence on wavelength also
becomes a lot weaker for larger temperature contrast \citep[see][for
more details]{2010ApJ...710..432R}.

\begin{figure}
  \centering \mbox{
    \includegraphics[width=.35\textwidth]{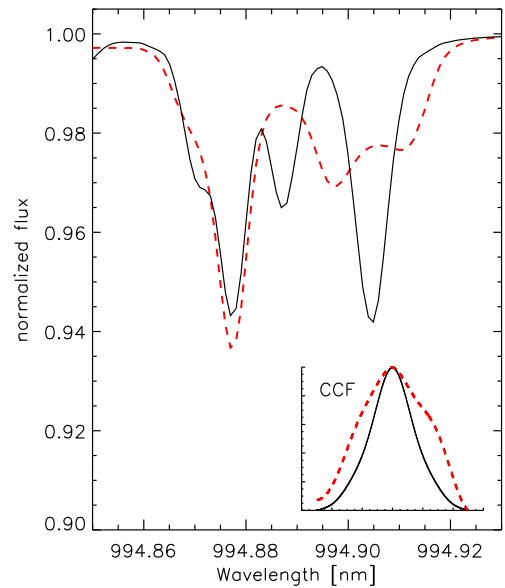}}
  \caption{\label{fig:FeHlines}Example of apparent velocity shift in
    magnetically sensitive spectral lines. Black solid line: spectrum
    containing four lines of molecular FeH without magnetic fields in
    a $T = 3700$\,K atmosphere; red dashed line: same spectral lines
    influenced by a magnetic field of $B = 1000\,$G. Inset shows
    auto-correlation function of the field-free case (black solid
    line) and cross-correlation between field-free spectrum with
    magnetic case spectrum (red dashed line).}
\end{figure}

\section{Symmetric line broadening}
\label{sect:symmetric}

\begin{figure*}
  \parbox{\textwidth}{
    \centering
    \mbox{
      \qquad
      \includegraphics[width=.32\textwidth,bb=100 0 618 438]{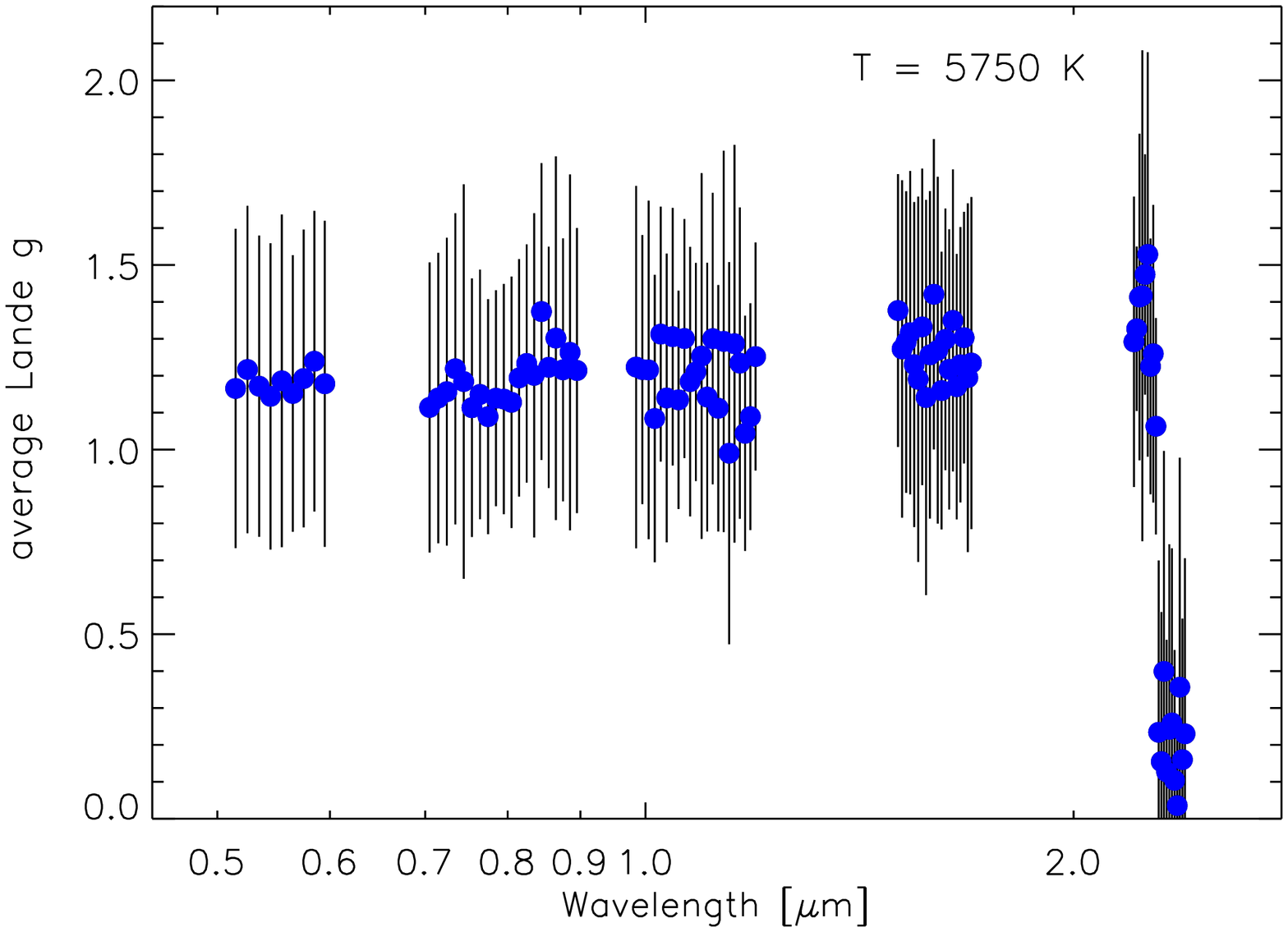}%
      \includegraphics[width=.32\textwidth,bb=100 0 618 438,clip]{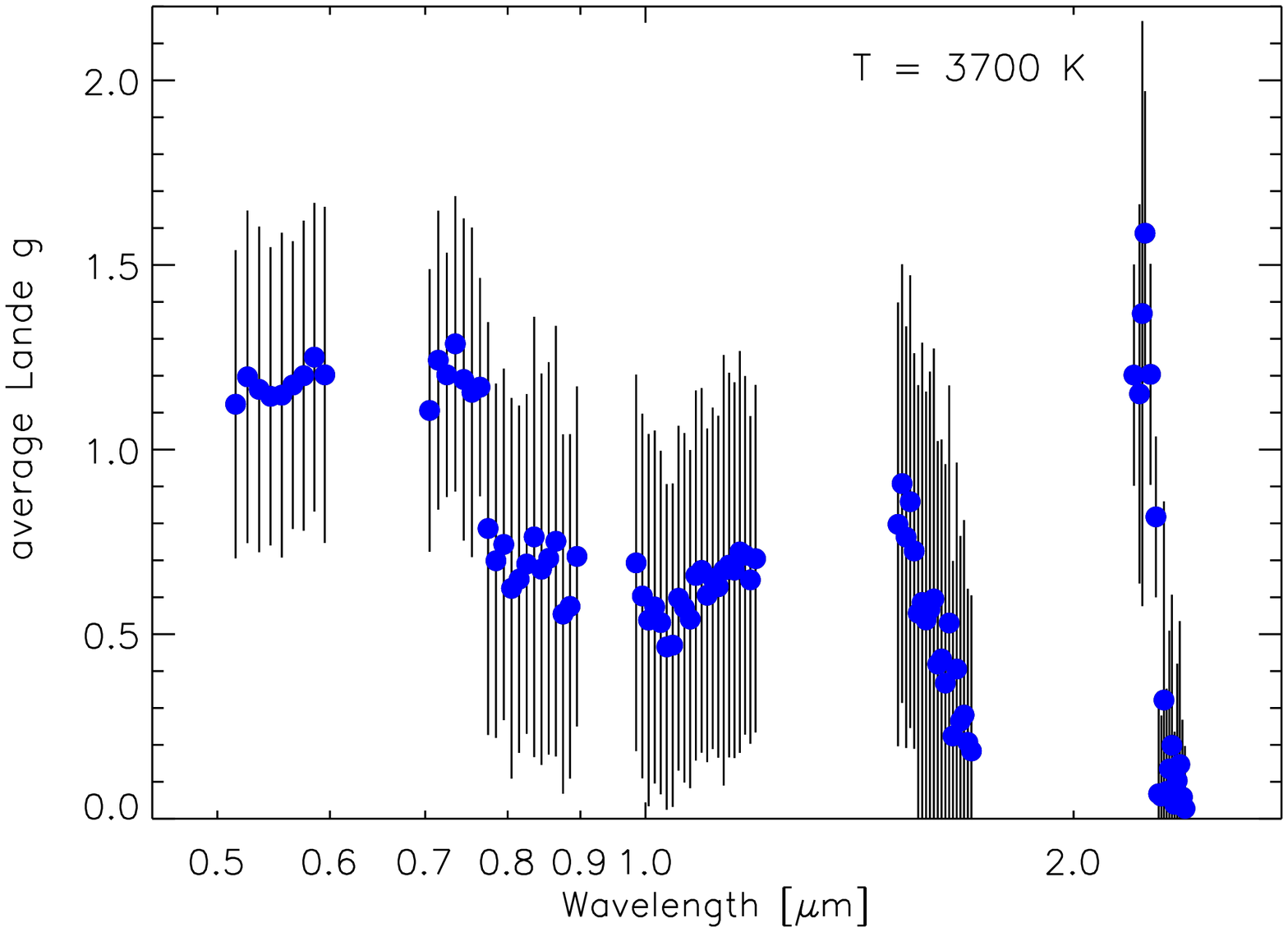}%
      \includegraphics[width=.32\textwidth,bb=100 0 618 438,clip]{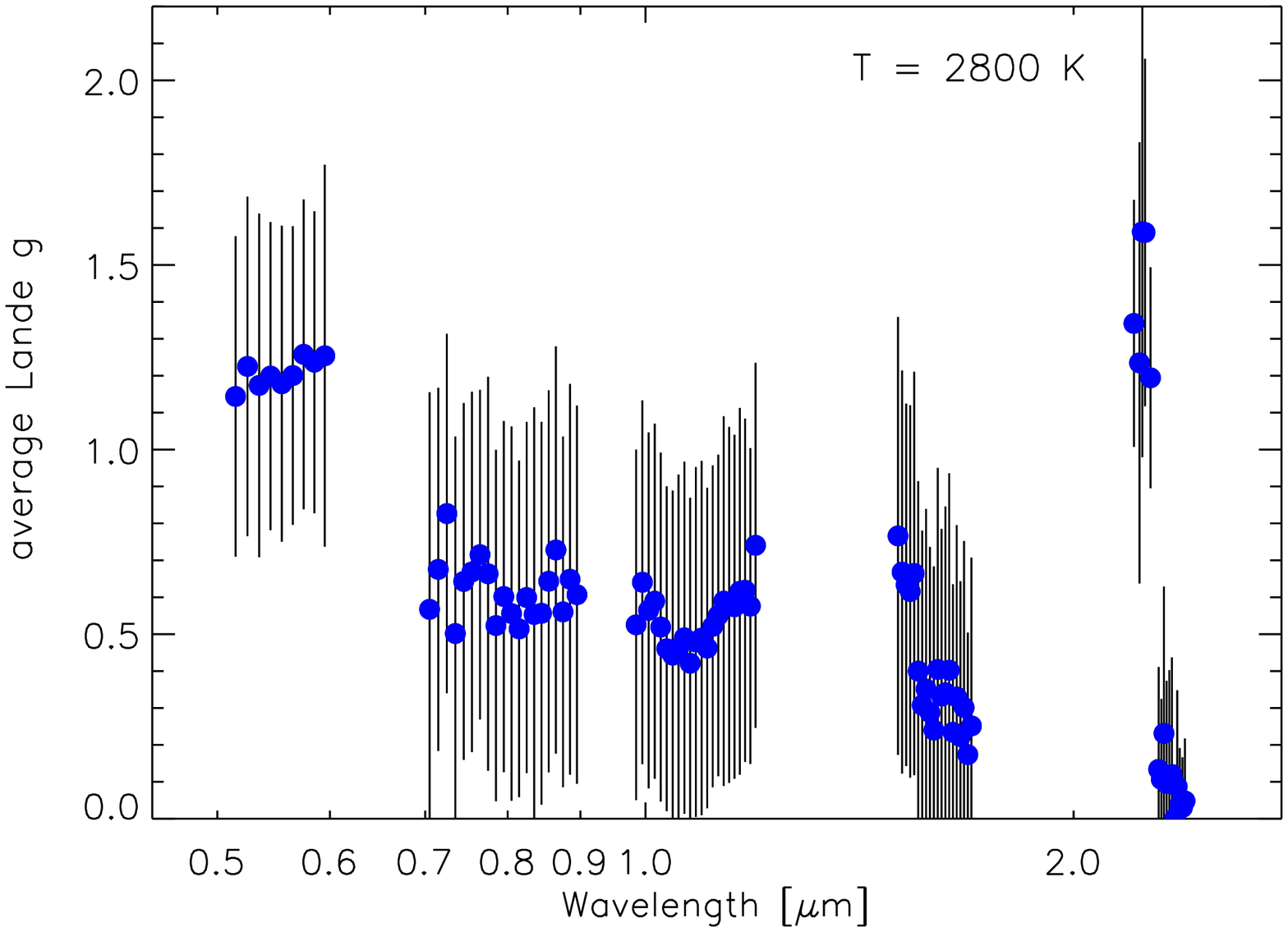}
    }}
  \caption{\label{fig:Lande}Average Land\'e g factors for lines deeper
    than 2\,\% of the continuum in the modeled spectra range (blue
    circles) together with rms scatter of all Land\'e g values
    considered in that region (error bars). The three panels show
    three different atmospheric temperatures as indicated in the
    figure. The low values at 2.3\,$\mu$m are due to the CO band.}
\end{figure*}

In addition to the signature of co-rotating spots, the effect of
variable average magnetic field distributed over the entire star can
also be very interesting. This could, for example, approximate the
effect of a magnetic cycle in an active star, or it can be caused by
stellar rotation since the average field visible at a given epoch can
differ from the one seen in other snapshots while the field
distribution is rather symmetric. However, in our model description of
the atomic and molecular Zeeman effect the pattern of Zeeman splitting
is always symmetric (see Section\,\ref{sect:Modeling}). Thus, for a
geometrically symmetric field distribution and single lines, no radial
velocity shift can be introduced because the spectral line's
barycenter always remains constant. On the other hand, the symmetry of
the appearance of several lines is broken as soon as line blending
occurs. This can be seen in Fig.\,\ref{fig:FeHlines}.  Here, the
difference between the two spectra is due to a change in the average
field not restricted to a starspot. Due to line blending, the two
spectra are now very different, and the barycenter of the
merit-function (the cross-correlation function or the goodness-of-fit
estimate) will be distorted introducing a spurious line
shift\footnote{The amplitude of this shift will in fact depend on the
  method the barycenter is located: fitting the dashed line in the
  inset in Fig.\,\ref{fig:FeHlines} will provide a different result
  than searching for the absolute maximum of that
  function.}. Deviations from the center of the non-distorted line may
be significant on the m\,s$^{-1}$ level, which is much less than
1/1000 of the line width.

Sign and amplitude of the RV shift depends on the sensitivity of the
lines to the Zeeman effect but also on the amount of blending between
different lines. Therefore, the apparent RV shift likely scatters
between different wavelength bands and the net result is probably
enhanced (random) jitter in the RV signal instead of a systematic RV
offset like in the case of co-rotating spots. We calculate this effect
and limitations predicting its amplitude in
Sect.~\ref{sect:avefieldresults}. To do so, we require accurate
spectral line data and a realistic description of the Zeeman effect
over large ranges of the stellar spectrum. We describe the line data
we use for our more sophisticated model in the next Section.

\section{Model atmospheres and line data}
\label{sect:Modeling}

\def\mum{$\mu$m}
\def\teff{T_{\rm eff}}
\def\logg{\log(g)}

For the detailed calculation of stellar spectral lines over large
wavelength regions, we used the
MARCS\footnote{http://marcs.astro.uu.se} model atmospheres
\citep{2008A&A...486..951G} using three different models:
$\teff=5750$~K, $\logg=4.5$ (a solar-type star), $\teff=3700$~K,
$\logg=5.0$ (an early-M-type star of spectral type $\sim$M1), and
$\teff=2800$~K, $\logg=5.0$ ($\sim$M6.5 star). For all models, solar
abundances according to \citet{2005ASPC..336...25A} were assumed. Our
theoretical spectra were computed using the {\sc synmast} code
\citep{2007pms..conf..109K} that can treat atomic and molecular
transitions in a magnetized medium.

Atomic line data were extracted from the VALD database
\citep{1995A&AS..112..525P, 1999A&AS..138..119K}. The Land\'e
g-factors were taken as provided by VALD or were computed using
available term designations assuming the LS-coupling approximation.

The spectra of cool stars are dominated by molecular
absorption. However, for most molecular lines the Zeeman effect is
only poorly understood.  The lack of laboratory measurements of
Land\'e g-factors and the complex physics behind molecular line
formation in plasmas with strong magnetic fields make it very
challenging to accurately model spectra of these objects. For this
work, we concentrate mostly on the effect of molecular FeH. We tested
theoretically computed Zeeman patterns on observations of a number of
M-dwarfs as described by \citet{2010A&A...523A..37S}. FeH line data
was taken from
\citet{2003ApJ...594..651D}\footnote{http://bernath.uwaterloo.ca/FeH/},
and we used corrected line intensities and positions following
\citet{2010A&A...523A..58W}. We also included FeH lines in the range
1.0--1.7\mum\ using the same procedure for computing Land\'e g-factors
as in \citet{2010A&A...523A..37S}. In addition to FeH lines, we also
included the line list of $X ^1\Sigma^+$ CO transitions from
\citet{1994ApJS...95..535G}. The well-known 2.3\mum\ band of CO is
often used for RV measurements because of its magnetic insensitivity
\citep[see, e.g.][]{2010ApJ...713..410B}.

We did not consider Zeeman splitting of molecular bands other than
from FeH and CO. Thus, the present investigation is only an
approximation of the effect of Zeeman splitting on RV signals; a
complete model should include all molecular lines present in the
stellar spectra. Nevertheless, FeH is the most important opacity
source in near-infrared spectra of very cool stars, and we believe
that the main effects from Zeeman splitting can be captured by our
approach. Prominent molecular bands not included in our model are due
to, e.g., TiO, CH, OH, and MgH. Some of them are known to exhibit
moderate or strong magnetic sensitivity as discussed in
\citet{2002A&A...385..701B}. Since our main goal is to characterize
the general behavior of RV signals comparing different wavelength
regions, and since a large part of the trends can be described
neglecting detailed line list information (see
Sect.\,\ref{sect:results}), we do not see a reason why the addition of
more molecular species should systematically change our
results. Nevertheless, quantitative predictions about absolute RV
distortions need to be interpreted with great care bearing in mind the
limits of our modeling approach. Including more magnetically sensitive
lines will also result in stronger blending, hence implications from
line blending presented in this work are probably lower limits.

Fig. \ref{fig:Lande} shows a compilation of average Land\'e g values
used in our model spectra. For parts of 100\,nm length, we calculated
average Land\'e $g$ values taking into account all atomic and
molecular lines stronger than 2\,\% (neglecting line broadening due to
surface motion and stellar rotation). The figure also visualizes the
spectral regions we considered for our investigation, which are
similar to the photometric bands $V, I, Y, J, H$, and $K$. Cooler
stars show more molecular absorption than hotter stars do. This
becomes apparent in the lower values of average Land\'e $g$ values
since on average molecular lines have lower $g$.

\section{Results}
\label{sect:results}

\begin{figure}
  \centering 
    \centering
    \includegraphics[width=.415\textwidth,bb=0 62.275 645 465,clip]{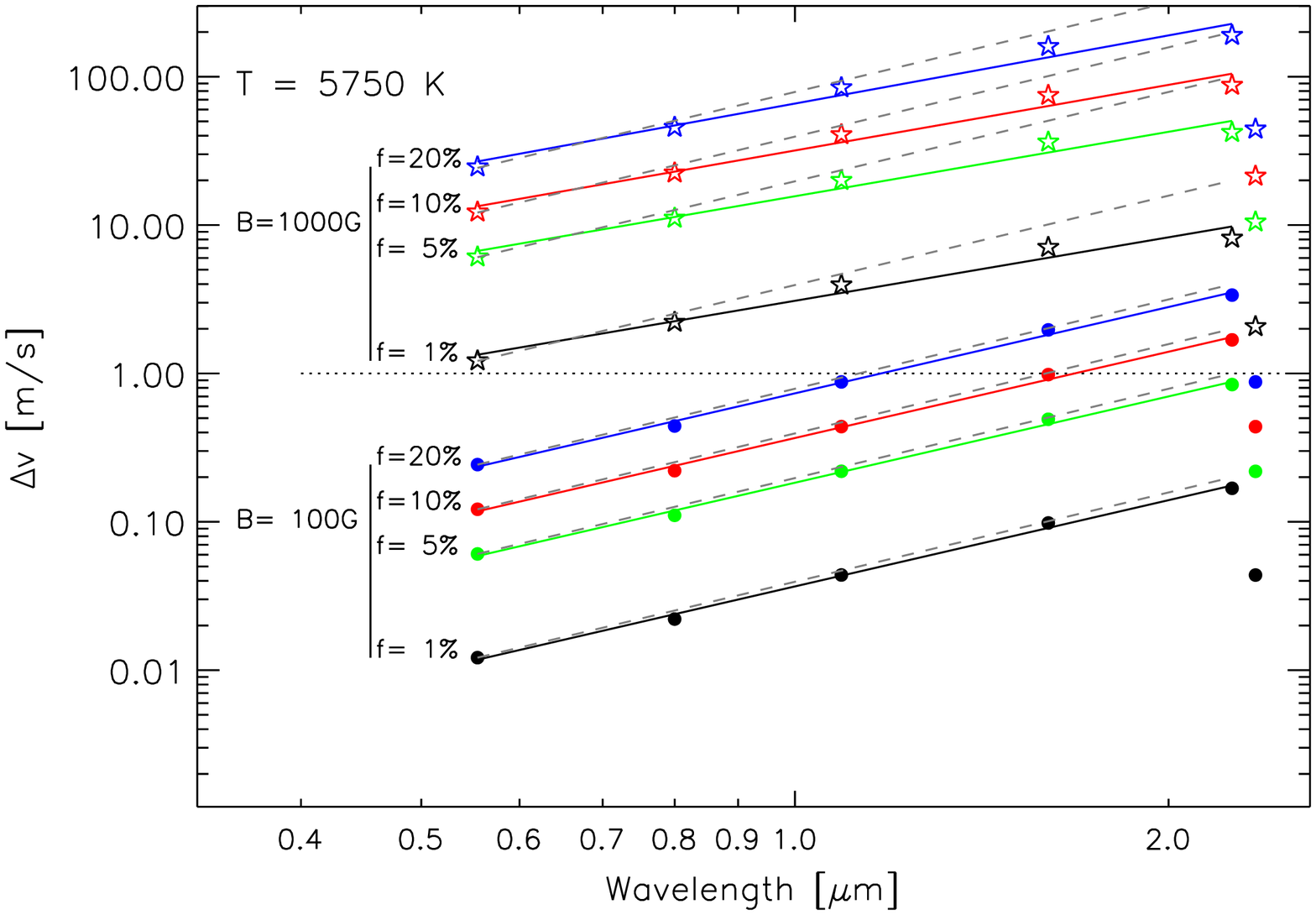}\\[-4mm]
    \includegraphics[width=.415\textwidth,bb=0 62.275 645 465,clip]{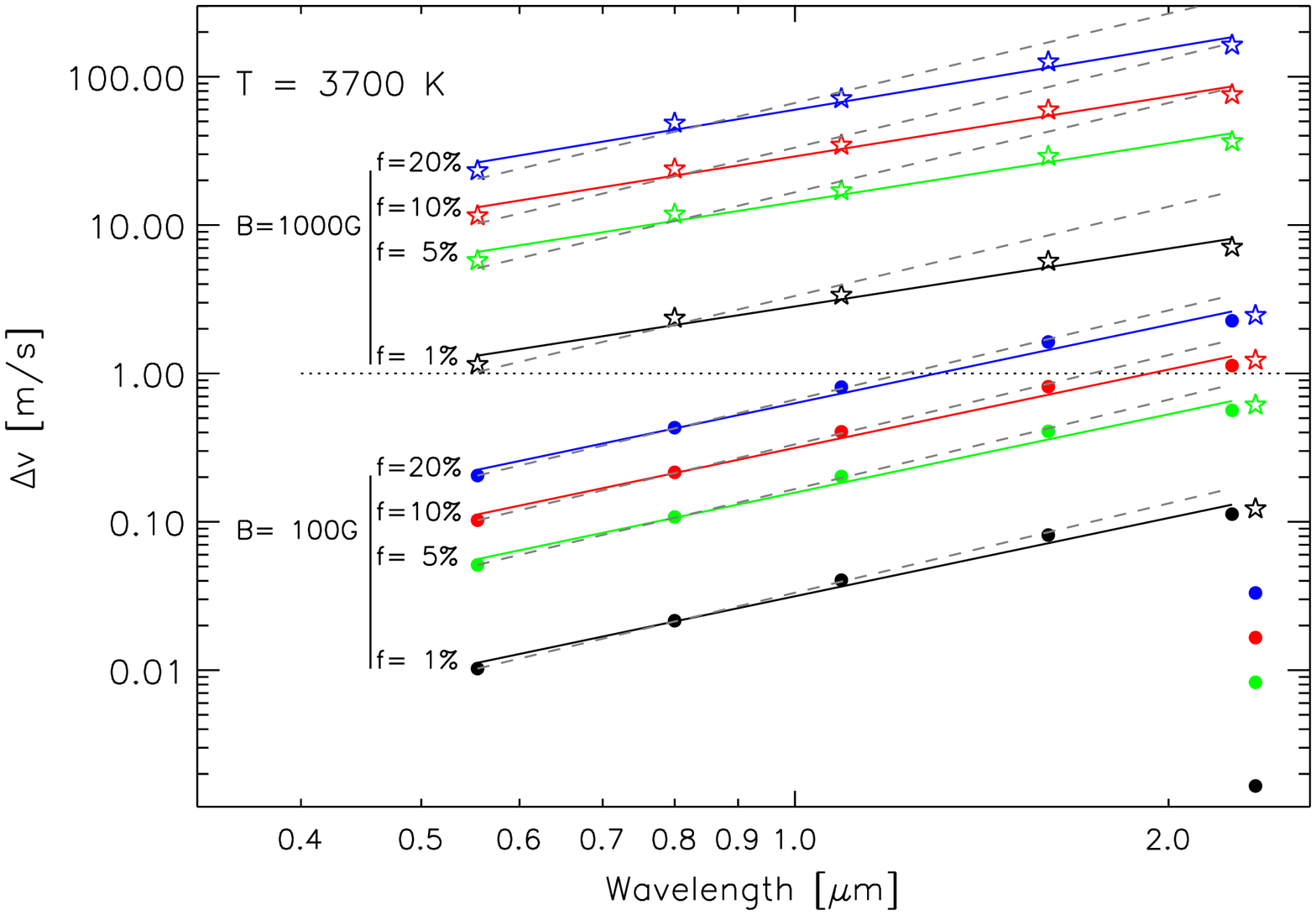}\\[-4mm]
    \includegraphics[width=.415\textwidth,bb=0 0 645 465]{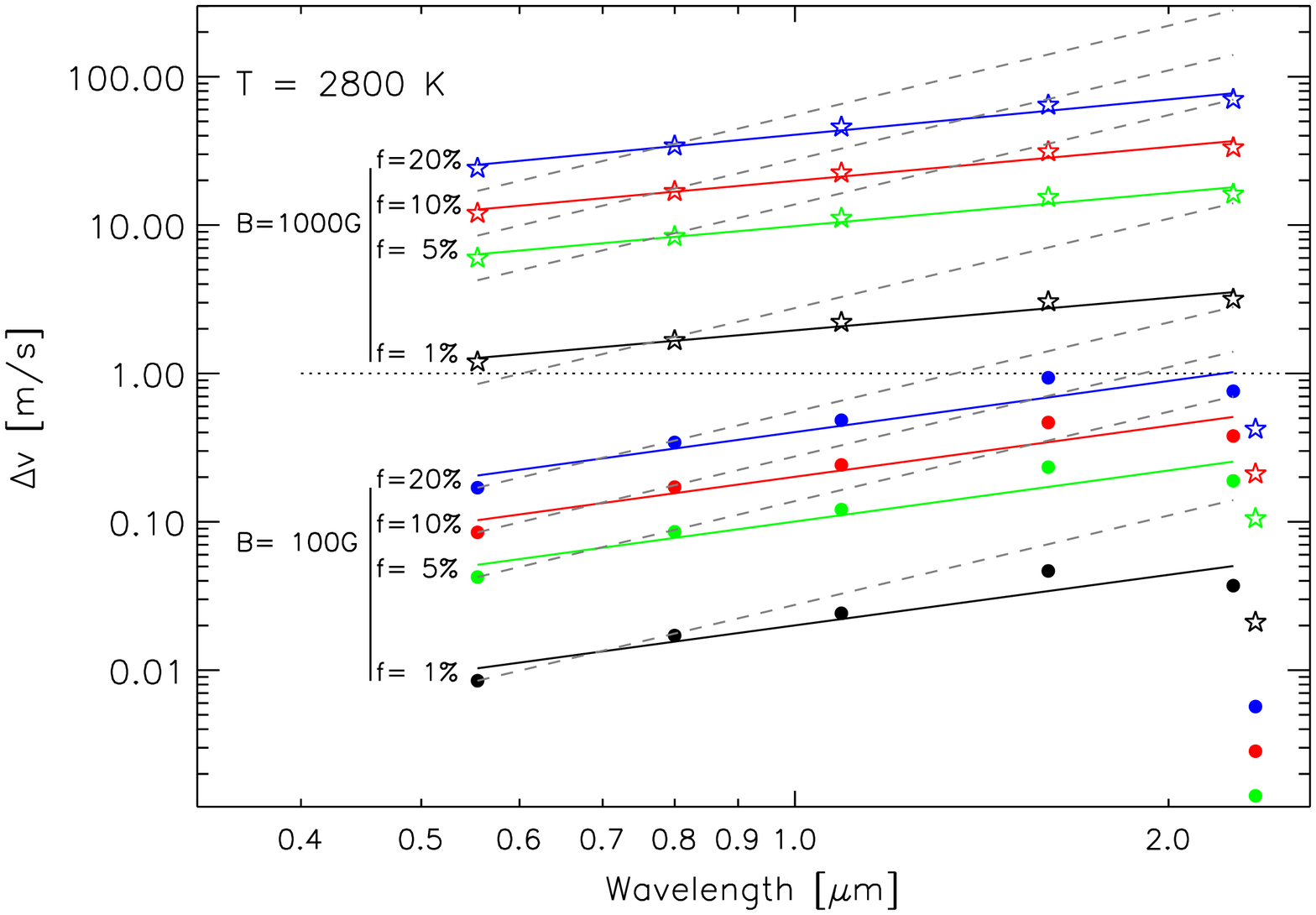}
  \caption{\label{fig:RVcorotation}Zeeman RV signature of co-rotating
    magnetic spots calculated from polarized radiative transfer over
    several wavelength bands. Stars and filled circles indicate six
    wavelength bands where the RV signature was measured, a fit to the
    first five bands using a straight line is shown as a solid line
    for each case (the CO band is not included in the fit). Dashed
    lines show the scaling expected from our our toy model, i.e.,
    $a=2$ in Eq.\,\ref{eq:realmodel}.}
\end{figure}

The mechanism causing an RV signal in an idealized spectral line
through the Zeeman effect was demonstrated in
Sect.\,\ref{sect:Toy}. The amplitude of an RV signal occurring in a
realistic spectrum will also depend on blending and the properties of
lines contributing to radial velocity information, foremost their
Land\'e factors and central wavelength. The influence on radial
velocity measurements in real stars can therefore be expected to
differ significantly from the results of our simple toy
model. Furthermore, radial velocity shifts from temperature contrast
and from Zeeman splitting lead to amplitudes that are comparable in
absolute values if calculated independently, but active regions on the
Sun are magnetic \emph{and} differ in temperature. In analogy to the
solar case, we expect that the stronger magnetic fields on other stars
are present in spots that are both cool and magnetic, but it is very
difficult to assess realistic values for temperature and magnetic
field contrasts. In the following, we first calculate the radial
velocity signature of a magnetic spot using not a single line but
synthetic spectra including several thousand atomic and molecular
lines that are split due to the Zeeman effect. After that, we show a
few example cases for magnetic, cool spots and the net radial velocity
signal from the two competing effects, and we calculate the RV signal
of average magnetic field variations using synthetic model spectra.

\subsection{Co-rotating magnetic spot}

We calculated model spectra of rotating stars with magnetic spots as
in our toy model above, but for the more realistic case we used
spectra from our polarized radiative transfer code. As above, the
cross-correlation function is calculated using the non-spotted
(non-magnetic) star as a template and the star with a magnetic spot as
our data set. For each case, we calculated the barycenter of the
cross-correlation function in six wavelength areas of more than
100\,nm each (see Fig.\,\ref{fig:Lande}). Three stars of temperature
$T = 5750$\,K, $T = 3700$\,K, and $T = 2800$\,K are calculated, spot
sizes of $f = 1\,\%, 5\,\%, 10\,\%$, and $20\,\%$, and field strengths
inside the spot of $B = 100$\,G and 1000\,G are used. The results are
shown in Fig.\,\ref{fig:RVcorotation}.

For our hottest example, the results from the synthetic model do not
differ very much from the trends we found in our toy model using a
single line with constant Land\'e g. In all cases, spots with magnetic
fields of 1\,kG introduce RV signals exceeding 1\,m\,s$^{-1}$ at all
wavelengths. The scaling of Eq.\,\ref{eq:toyscaling} is a very good
approximation to the situation seen in the case for $T = 5700$\,K. In
our cooler examples, however, the values we calculate from the
synthetic atmosphere models show significantly less dependence on
$\lambda$ in the sense that the radial velocity signal does not grow
proportional to $\lambda$ squared. Nevertheless, we can approximate
the radial velocity signal in all three examples using the following
formula:

\begin{equation}
  \label{eq:realmodel}
  \Delta v_{\rm{rad}} = 300 \frac{\rm m}{\rm s} \,f \, \, \left(\frac{B}{\small{\textrm{kG}}}\right)^2 \left(\frac{\lambda}{\small{\mu\textrm{m}}}\right)^{a},
\end{equation}
with the relative fraction of the spot area, $f$, the magnetic field
inside the spot, $B$, and the wavelength $\lambda$. The value of the
constant results from the geometry of the surface and on the
distribution of Land\'e factors across the wavelength range, but not
on other free parameters. While $a \approx 2$ in our hottest example
(same value as in our toy model), we find $a<2$ for cooler stars. The
reason for this is clear from the distribution of Land\'e g-values
shown in Fig.\,\ref{fig:Lande}. While in the sun-like case ($T =
5750$\,K) the typical Land\'e g-values are not a function of
wavelength (the CO band being the only exception), Land\'e g-values
are significantly lower at longer wavelengths in the cooler stars,
partially compensating for the linear increase of the total Zeeman
shift as a function of wavelength (see Eq.\,\ref{eq:deltav}). This is
because at cooler temperatures, molecular lines become more and more
important in relation to atomic lines, and (at least in our model) the
molecular lines tend to have lower Land\'e g-values on average.

A central result of our simulation is that the radial velocity signal
due to a co-rotating spot of 1\,\% the size of the projected stellar
disk and a magnetic field strength of 1\,kG has an amplitude of
approximately 3\,m\,s$^{-1}$ if observed at $\lambda = 1\,\mu$m. The
amplitude scales linearly with filling factor and quadratically with
both magnetic field strength and wavelength. In stars significantly
cooler than the Sun, the scaling with wavelength is weaker than
quadratic because the relevant spectral features are magnetically less
sensitive.

\subsection{Cool and magnetic spot}

\begin{figure}
  \centering \mbox{
    \includegraphics[width=.48\textwidth,bb=30 65 450 590,clip]{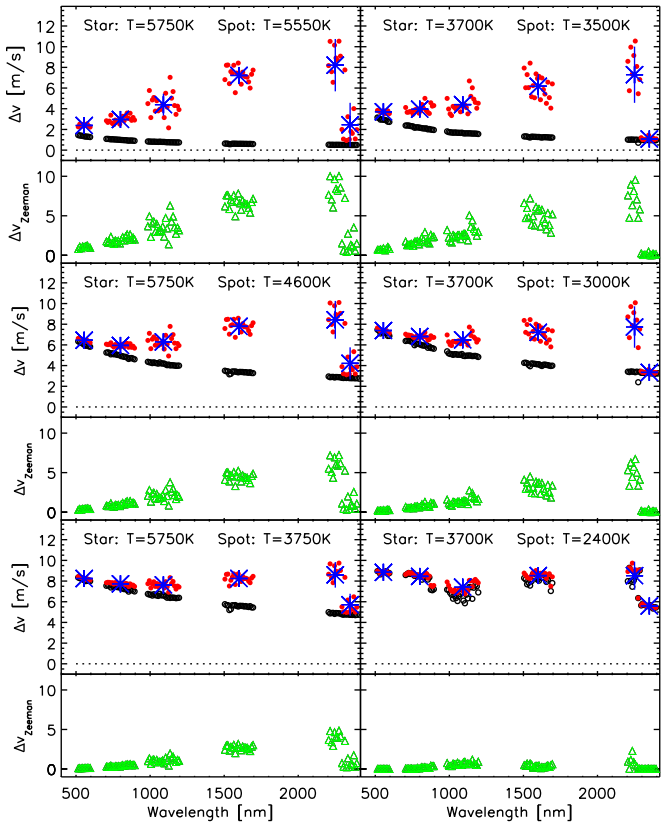}}
  \caption{\label{fig:coolspots}Radial velocity amplitude due to
    co-rotating cool, magnetic spots ($f = 1$\,\%, $B = 1000$\,G,
    $v\,\sin{i} = 2$\,km\,s$^{-1}$) using our synthetic atmosphere
    model for different star/spot temperatures. Black circles show the
    influence of temperature only ($B = 0$), red circles show the net
    effect including temperature contrast and Zeeman broadening in the
    spot. Blue crosses are average values and range for six wavelength
    bands. Green triangles show residuals between red and black
    circles, i.e., the effect due to Zeeman broadening. Left panel
    shows a sun-like star with $T = 5750\,$K, right panel shows an
    early M-type star with $T = 3700$\,K. Top to bottom panels show
    different values of $\Delta T = T_{\rm star} - T_{\rm spot}$; top
    panel: $\Delta T = 200$\,K; middle panel: $\Delta T \approx T/5$;
    bottom panel: $\Delta T \approx T/3$.}
\end{figure}

In the preceeding sections and in \cite{2010ApJ...710..432R}, the
effects of magnetic spots (at the same temperature as the photosphere)
and of (non-magnetic) temperature spots were modeled independently. In
reality, of course, spots are understood to be both magnetic and
effectively cooler than the quiet photosphere. As a first
approximation, we compared both effects. We found that temperature and
magnetic effects can cause radial velocity signatures of comparable
amplitude, but temperature effects are larger at short wavelengths
while magnetic influence is more significant at long wavelengths.

The first order approximation looking at both effects independently is
probably not realistic because a low temperature spot emits less flux
than a hotter one, which will lead to a weaker influence of the Zeeman
effect on the line profile and its dependence on wavelength. In order
to consider the two effects in a consistent way, we calculated models
of stars with spots that are both cool and magnetic, and we show the
results of radial velocity amplitude in Fig.\,\ref{fig:coolspots}. In
all cases, spot parameters are $f = 1\,\%$ and $B = 1000$\,G, the star
is assumed to be rotating at $v\,\sin{i} = 2$\,km\,s$^{-1}$. We show
results for a sun-like star with $T = 5750$\,K and for a cooler
(M-type) star with $T = 3700$\,K; the $T = 2800$\,K does not provide
new information and we do not include it in this example. For spot
temperatures, three cases are considered for each star: one with
$\Delta T = T_{\rm star} - T_{\rm spot} = 200$\,K (top panel), one
with $\Delta T \approx T/5$ (middle panel), and a third with $\Delta T
\approx T/3$ (bottom panel).

Our first result is that in all cases the net effect is a non-trivial
combination of temperature contrast and magnetic Zeeman
splitting. Both mechanisms work in the same direction; the total RV
signal accounting for both effects is always larger than the signal
from one mechanism alone. The net effect reaches values up to the
10\,m\,s$^{-1}$ level. In our examples with the lowest spot/star
temperature ratio ($\Delta T = 200$\,K), the radial velocity signal
monotonically grows with wavelength. In our intermediate cases of
temperature contrast ($\Delta T \approx T/5$, center row), the net
signal has a local minimum around $1\,\mu$m, it is dominated by
temperature contrast effects at shorter wavelengths and by Zeeman
splitting at longer wavelengths. In our examples with the highest
temperature contrast (bottom panel), the wavelength dependence of the
total RV signal is nearly constant with wavelength: in the sun-like
star (bottom left), the RV signals from the temperature effect and
from the Zeeman effect almost cancel; in the cooler star with the
coolest spot (bottom right), the RV signal is always dominated by the
temperature contrast. In the latter case, the spot contributes so
little flux that the Zeeman signal is not significant even at the
longest wavelengths used here. Interestingly, the signal from the
temperature spot alone shows some scatter with wavelength that is not
monotonic in $\lambda$ and is produced by the temperature dependence
of individual absorption lines that show different intensities inside
and outside the spot region \citep[cp.][]{2010ApJ...710..432R}. For
example, the CO lines in this case become deeper with lower
temperature, which counteracts missing line absorption emerging from
the spot area and leads to a reduced RV signal within the CO line
region.

\begin{figure*}
  \centering \mbox{
    \includegraphics[width=.49\textwidth,bb=00 90 620 340]{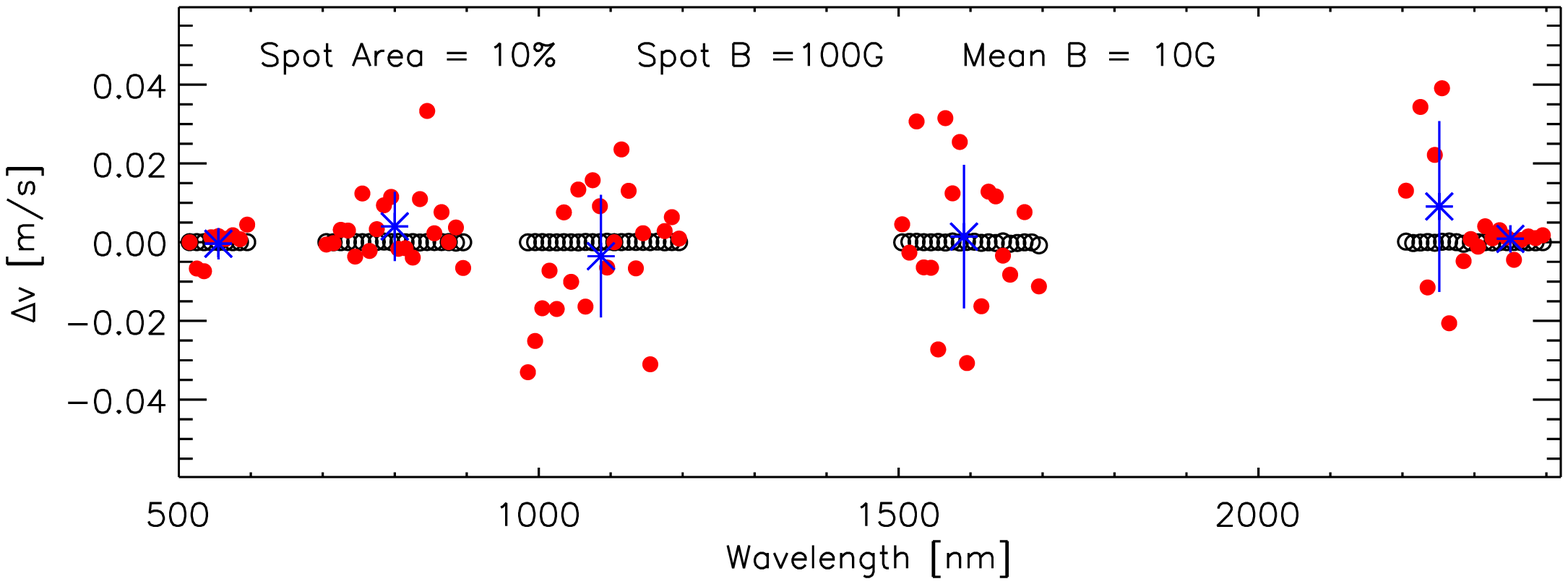}
    \includegraphics[width=.49\textwidth,bb=00 90 620 340]{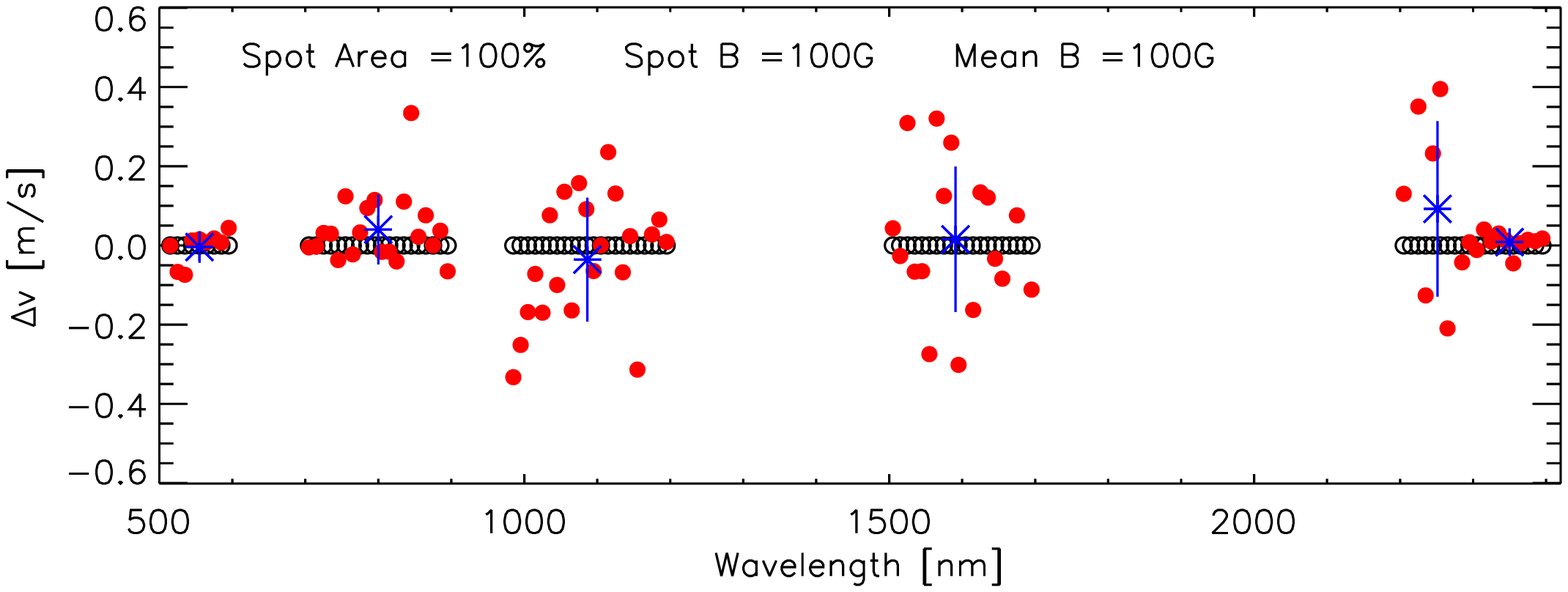}}
  \caption{\label{fig:diffuse}Radial velocity scatter due to variable
    average magnetic field. Red circles show results from
    cross-correlation between spectra of a non-magnetic star and a
    star with an average magnetic field. Blue crosses and error bars
    show average and rms-scatter of individual chunks for six
    wavelength bands. Left panel: surface fraction $f = $10\,\% of the
    star is covered with magnetic field varying between 0 and 100\,G;
    right panel: the entire star ($f = $100\,\%) shows magnetic field
    variability of 100\,G. The small scatter at 2.3\,$\mu$m is due to
    the CO-band. Note the different scaling on the y-axis.}
\end{figure*}

Even if the model we show in Fig.\,\ref{fig:coolspots} includes
radiative transfer of atomic and molecular spectral lines, and both
the influence of temperature and Zeeman splitting, we are aware of the
problem that our model is probably very different from any real
star. Values of spot temperature and field strengths in our model are
probably not unrealistic for some individual spots, but the real range
in temperature and field strengths are largely unkown. Perhaps more
important, active stars most probably are very different to a one-spot
model and evolve in time. Therefore, we restrict ourselves to the
examples shown in this section and do not attempt to make any more
specific predictions on radial velocity signals in active stars.

We conclude from our examples that in active stars, both the influence
of temperature contrast and the Zeeman effect can be of comparable
amplitude, and that the details of spot distribution and their
temperatures and magnetic fields determine amplitude and
wavelength-scaling of the RV signature due to activity. Thus, radial
velocity signals due to active regions cannot be expected to vanish at
infrared wavelengths. On the contrary, in many stars the influence of
starspots may be much more severe at longer wavelengths. The best way
to discriminate between a planet and a starspot is therefore
simultaneous measurement of radial velocities at many different
wavelengths; even if the scaling of the radial velocities with
wavelengths is difficult to predict, it is improbable that a signal
due to co-rotating active regions is independent of wavelength. Any
wavelength-dependent signal will rule out companions as source, and
the scaling with wavelengths will provide useful information on the
nature of active regions.

\subsection{Average field variations}
\label{sect:avefieldresults}

\begin{figure}
  \centering \mbox{
    \includegraphics[width=.475\textwidth]{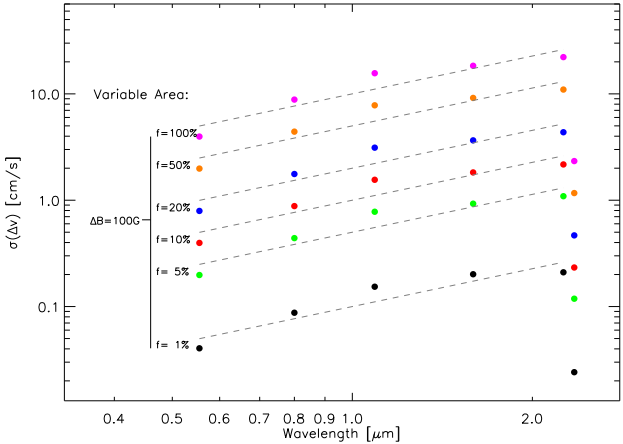}}
  \caption{\label{fig:RVdiffuse}Zeeman-induced RV scatter for the case
    of variable average magnetic fields. Six cases are shown in which
    the fractional coverage of the field is between 1\,\% and
    100\,\%. In that area, the field varies by 100\,G. The scatter
    depends on wavelength and can be approximated by the dashed grey
    lines calculated from Eq.\,\ref{eq:averagefield}.}
\end{figure}

Magnetic and cool spots on the surface of a rotating star introduce
line profile variations as discussed above. Radial velocity signals
due to this effect are introduced by the deformation of all individual
spectral lines and show the same period as the rotation of the
star. Another effect caused by the Zeeman effect was introduced in
Section\,\ref{sect:symmetric} and is due to systematic mismatch
between large spectral regions observed at one time with respect to
another observation. A possible reason for such a mismatch can be
variation in the (average) magnetic field of a star, for example
during a magnetic cycle. In a single line, average-field variability
would not lead to a radial velocity signal assuming symmetric Zeeman
splitting (as always assumed here). In spectral regions containing
many lines, however, an apparent shift may be introduced because lines
are usually blended with others (see Fig.\,\ref{fig:FeHlines}).

In Fig.\,\ref{fig:diffuse}, we show the results from cross-correlating
a spectrum with no magnetic field with a spectrum of average field of
$Bf = 10$\,G (left panel) and $Bf = 100$\,G (right panel). For field
variations with respect to a non-zero field we expect similar results
since Zeeman splitting is linear in B. The spectra are constructed
assuming $B = 100$\,G field strength in active regions homogeneously
covering $f = 10$\,\% (10\,G case) and $f = $100\,\% (100\,G case) of
the projected stellar disk. The examples represent the cases in which
the average magnetic field varies by $\Delta B = 10$\,G and $\Delta B
= 100$\,G, respectively.  We show results from individual wavelength
regions covering 10\,nm each. In contrast to the cases with
co-rotating spots, the radial velocity shift does not follow a
systematic pattern because the signal we measure here is a result of
the random blending of broadened lines. On average, the signal is zero
but variations in the field strength introduce scatter that depends on
the properties of the field variations and wavelength. In
Fig.\,\ref{fig:diffuse}, we show average values (blue crosses) and
rms-scatter for wavelength bands containing several adjacent
individual wavelength parts. As expected, we find that the average
values of the RV shift is consistent with zero at all bands, but the
rms grows with wavelength (an exception is again the CO band at
2.3\,$\mu$m).

In Fig.\,\ref{fig:RVdiffuse}, the rms-values of the RV signal are
displayed as a function of wavelength for the two models of
Fig.\,\ref{fig:diffuse} plus four more models with fractional coverage
between $f = $1\,\% and 100\,\%. Note that the amplitude of rms
scatter is given in cm\,s\,$^{-1}$. We fit a power law to our results
(grey dashed lines in Fig.\,\ref{fig:RVdiffuse}) and find an
approximation to the RV scatter introduced by variations in the
average magnetic field:

\begin{equation}
  \label{eq:averagefield}
  \sigma(v_{\rm{rad}}) = 10 \frac{\rm cm}{\rm s} \, \,f \, \left(\frac{\lambda}{\small{\mu\textrm{m}}}\right)^{1.18},
\end{equation}
for $\Delta B = 100$\,G. This approximation is valid for all
wavelength bands except the one containing the CO lines. For the
$\Delta B = 10$\,G case, we find that while the effect is of course
smaller than the one for $\Delta B = 100$\,G, there is no simple
scaling relation that relates the cases of different $\Delta
B$. Nevertheless, we can conclude that additional scatter in radial
velocity measurements can be introduced by variable average magnetic
fields. The uncertainty of a radial velocity measurement at a given
wavelength can be affected by field variability, but the amplitude of
this effect is likely well below 1\,m\,s$^{-1}$ in realistic
cases. Even stars with very strong magnetic fields are not believed to
show variations in average field strength on the order of kilo-Gauss,
and such stars would probably show very strong variability in
chromospheric emission lines, too, which could help to identify such
cases.

\section{An active star example: AD Leo}

Our calculations predict that magnetically active stars may show
wavelength-dependent RV variations with a larger RV signal at longer
wavelengths. For a first test of our results, we searched the
HARPS-ESO archive for magnetically active stars with clear
periodicities detected on them. A very clear and prominent example we
found was AD Leo (Gl~388, Spectral Type M4.5Ve). HARPS observations on
this star span a time baseline of 900\,d and have a typical SNR of
$50$ at 600\,nm. As reported by \citet{bonfils:2012}, AD~Leo shows a
very strong periodic signal in the RVs at 2.22\,d, which is consistent
with the stellar rotation period found in the Zeeman Doppler Imaging
analysis by \citet{morin:2008}. At the present time, 40 spectra are
publicly available. We analyzed the data using the HARPS-TERRA
software \citep{anglada:2012a}. HARPS-TERRA derives RV measurements by
least-squares matching each observation to a high signal-to-noise
(SNR) template generated by co-adding all available observations. It
also allows to obtain the Doppler measurement using a limited number
of echelle orders at a time, enabling the analysis of RV signals as a
function of wavelength. In order to accumulate enough SNR to derive
good quality RV measurements, we split the stellar spectrum in 7 parts
using 419, 450, 486, 528, 583, 631 and 665\,nm as the central
wavelength of each part. Except for the redder two parts, each part
spans ten HARPS Echelle orders, that is: orders 10--19, 20--29,
30--39, 50--59, 60--66, 68--71. The last two parts are chosen to avoid
order 67 containing H$\alpha$ that is highly variable in active stars.
In all these parts, the 2.22\,d period is clearly detected in the
periodograms \citep{cumming:2004}. The 10 bluest echelle apertures
(0--9) are not discussed here because the SNR is very low at the
bluest orders (at the 5th echelle aperture it was typically below 5),
and uncertainties associated to each individual RV were of the order
of 50\,m\,s$^{-1}$.

\begin{figure}
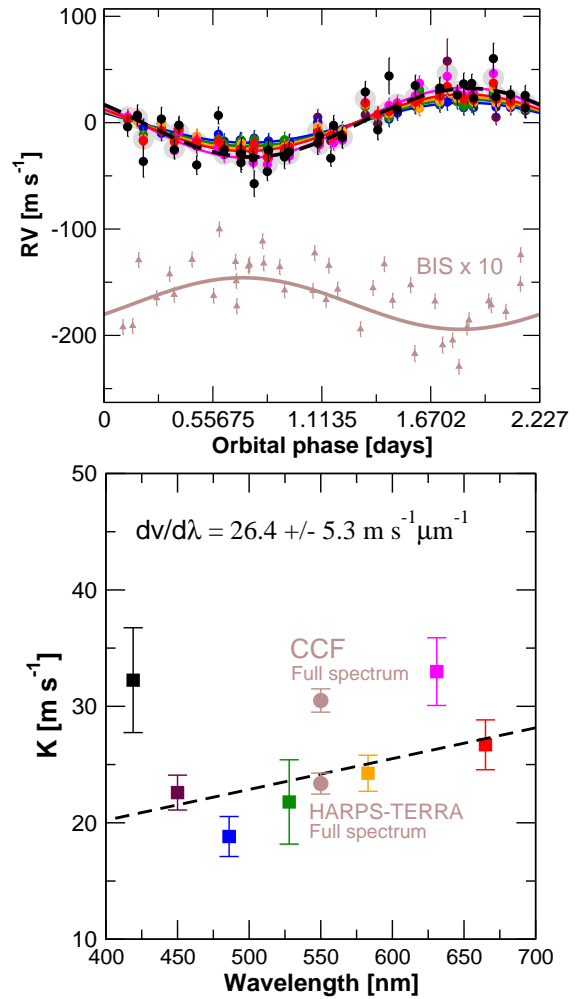

  \centering
  \includegraphics[width=0.4\textwidth,clip]{Phase_folded.eps}\\
  \includegraphics[width=0.38\textwidth,clip]{K_vs_Wave.eps}
  \caption{\label{fig:K_vs_wave}\emph{Upper panel:} Phase folded radial
    velocities and fitted signals to the preferred Doppler period of
    2.22704~d. The phase folded fit to the BIS is also provided. For
    illustration purposes the BIS data has been shifted and multiplied by
    10. \emph{Right panel:} Semi-amplitude $K$ of the signal as a function of
    wavelength. The best fit to a linear model representing the wavelength
    dependence of $K$ is given as a thick dashed line. The semi-amplitudes
    derived from the full spectrum (using CCF and HARPS-TERRA measurements)
    are also show as solid brown circles. Note that both measurements provide
    incompatible results providing a further test to assess the reality of a
    Doppler signal.}
\end{figure}

\begin{figure}
  \centering
  \includegraphics[width=0.45\textwidth,clip]{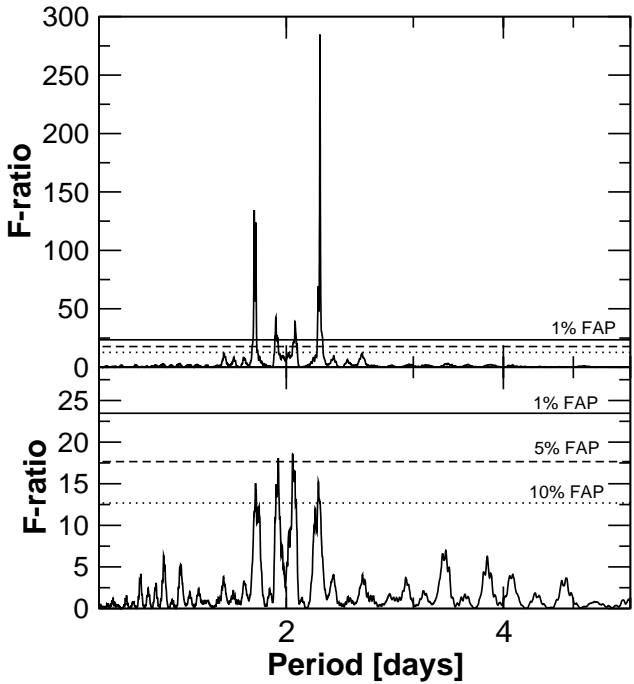}
  \caption{\label{fig:periodograms} Top: Periodogram of the radial
    velocities obtained using the full spectral range of
    HARPS. Bottom: Periodogram of the BIS. The 1\% FAP threshold is
    illustrated as solid horizontal lines in both panels. The four
    peaks in the bottom panel are likely strong aliases of the same
    signal (2.22 days would be compatible with the photometric period
    of the star). Note that their average FAP is only 5\%.  }
\end{figure}

In the upper panel of Fig.\,\ref{fig:K_vs_wave}, we show our fit with period
$P$=2.22704\,d to the RVs derived for each part. For simplicity, a sinusoid
with the fixed period derived from the best fit to the RVs using the full
spectrum (see top periodogram in Fig.\,\ref{fig:periodograms}) was adjusted to
each part, so the only free parameters were amplitude and phase of the
signal. In the lower panel of Fig.\,\ref{fig:K_vs_wave}, we plot the derived
semi-amplitudes as function of central wavelengths for each part. The
uncertainties were derived using the bootstrap technique, i.e., computing the
scatter of the amplitude as obtained by randomly selecting samples of the
observations. The HARPS-Data Reduction Software also provides a measure of the
mean spectral line asymmetry, called the bisector span (BIS). BIS is a measure
of the asymmetry of the cross-correlation function in RV space as obtained
from cross-correlating the stellar spectrum to a binary mask \citep[M2 binary
mask, see][]{pepe:2002}. As demonstrated in several studies
\citep[e.g.][]{queloz:2001}, BIS often anti-correlates with spurious RV
offsets if a cool spot is responsible for the apparent RV shifts. The BIS
periodogram of AD~Leo shows 4 peaks of similar power at 1.813, 1.950, 2.041,
and 2.219\,d (bottom panel in Fig.\,\ref{fig:periodograms}). These periods are
all related through daily aliases (standard and sidereal day) and they likely
correspond to the same physical periodicity. Since none of the BIS periods
exhibits a false alarm probability lower than 1\%, the periods are subject to
significant uncertainties; only an approximate match can be done when
comparing the BIS measurements with those in the RV signals.

The photometric period (and, presumably, the rotation period of AD
Leo) has been reported to be 2.23\,d \citep{2009AIPC.1135..221E}
favoring the period of 2.219\,d as the most likely fundamental signal
in the BIS. To see how it compares to the RV signal, we fixed the
period of a sinusoid to the preferred RV period and adjusted the
amplitude and phase of the BIS curve (see
Fig.\,\ref{fig:K_vs_wave}). Doing this we found that BIS appears to be
anti-correlated with the RV curve, which is consistent with the
expectations for a cool spot-induced signal.  Note that, for
visualization purposes, the BIS measurements in
Fig.\,\ref{fig:K_vs_wave} were multiplied by a factor of 10 (and are
offset). Also note that while the 2.22\,d periodicity is clearly
detected in the RVs, the F-ratios of the BIS candidate signals are 15
times smaller and barely significant (see
Fig.\,\ref{fig:periodograms}). We also looked at other activity
indicators typically associated to spurious RV signals on M dwarfs
\citep[e.g., FWHM of the CCF or the S-index; see][]{lovis:2011,
  anglada:2012b}, but we did not find any further indication of a peak
near 2.22\,d.

Assuming that the Doppler signal is caused by cool spots (neglecting Zeeman
broadening), one would expect its amplitude to become weaker at redder
wavelengths. We obtained a weighted fit to our values of $K$ shown in the
bottom panel of Fig.\,\ref{fig:K_vs_wave} using a linear model of the form
$K[\lambda] = A \lambda + b$, where A is the slope and is measured in units of
velocity per unit of wavelength (m\,s$^{-1}\,\mu$m$^{-1}$). The obtained slope
is 26.4\,m\,s$^{-1}\,\mu$\,m$^{-1}$ which is positive and different from 0 at
a 5-$\sigma$ level. Therefore, we conclude that the RV signal does not
diminish towards longer wavelengths in the range covered by HARPS.  Instead,
the obtained wavelength dependence of the amplitude has a significantly
positive slope with larger amplitudes $K$ towards longer wavelengths. Our data
do not extend into the infrared wavelength range, and we cannot reach any firm
conclusion on the behaviour of the RV signal wavelengths longer than
700\,nm. Nevertheless, our example provides first evidence that the RV signal
of an active star does not always diminish at larger wavelengths. Since RV
signals from cool spots due to temperature contrast alone are supposed to show
a monotonic behavior across optical and infrared wavelength regions, we
interpret this as evidence for another mechanism causing the positive slope of
the RV curve. The results of this paper would indicate that the Zeeman effect
is the reason for this, and that the amplitude of this signal would be even
larger at infrared wavelengths.

We note that the semi-amplitude derived from CCF measurements differs
from the one derived from HARPS-TERRA RVs using the full spectrum; the
result from CCF is $K = 30.5 \pm 1.0$\,m\,s$^{-1}$, the result from
HARPS-TERRA is $K = 23.4 \pm 0.9 $\,m\,s$^{-1}$. This indicates that
the changes in the line shapes affect each method in a very different
way, further indicating that the measured RV offsets are due to
changes in the line profile shapes rather than real Keplerian
signals. In conclusion, even in this example, in which we know the
photometric period and see an anti-correlation between the RV with
line asymmetries, the assumption that the RV signal is induced by the
temperature contrast effects alone is likely to produce an incorrect
interpretation of the data and can lead to the erroneous prediction
that the spurious RV signal is suppressed at nIR wavelengths. This
misconception would have serious consequences if one attempts to
correct the RV curve for activity signals. Nevertheless, it is
certainly true that Keplerian signals cannot be
wavelength-dependent. If stellar activity on a time-scale similar of a
Doppler signal is suspected, only a comprehensive analysis of its
wavelength dependence can shed light on its true physical origin.

\section{Discussion}

We have investigated the influence of magnetic activity on radial
velocity measurements in active stars. In contrast to earlier
calculations, we included the Zeeman effect and calculated line
barycenter shifts due to spots that are cool and
magnetic. Furthermore, we looked at the case of varying average
magnetic fields that are not concentrated in co-rotating regions and
may introduce signals not in phase with stellar rotation.

Our most important result is that co-rotating magnetic starspots can
be expected to significantly distort stellar line profiles and RV
measurements. Neglecting the cool temperature of the spots, the signal
from the Zeeman effect alone easily exceeds the 1\,m\,s$^{-1}$ level
even for very small active regions ($f = 1$\,\%) and slow rotation
($v\,\sin{i} = 2$\,km\,s$^{-1}$) if the field inside the spot is
comparable to sunspot fields ($B \sim 1$\,kG). This signal has the
same sign as the signal from a non-magnetic, cool spot, it grows with
both magnetic field and wavelength, and is approximately four times
larger at $\lambda = 1000$\,nm than at $\lambda = 500$\,nm. Its
amplitude saturates for very strong fields above approximately
1\,kG. Comparable radial velocity signals are also expected from the
temperature contrast of cool spots alone (neglecting the Zeeman
effect) as shown in earlier investigations, but they are largest at
short wavelengths. We note that the effect of Zeeman broadening on
integrated (non-polarized) light is independent of magnetic
polarity. Therefore, the RV signature does not depend on magnetic
polarity, and in particular does not cancel out if magnetic areas
consist of several spots with opposite polarity.

In contrast to the systematic fake RV signals from co-rotating spots,
the RV signal due to variable average (non-localized) fields is of
statistical nature only affecting the noise floor of the
measurements. The typical uncertainty of jitter induced by an average
magnetic field varying by 100\,G is on the 10\,cm\,s$^{-1}$ level. No
systematic signal is expected from variations in the mean magnetic
field strength.
 
Including both temperature contrast and the Zeeman effect is necessary
to understand RV signals in active stars. In case of large contrast
(very cool spots), the Zeeman effect is less important because the
integrated spectrum contains less flux from the magnetic (spot)
region. Nevertheless, since temperature contrast diminishes at long
wavelengths and the influence of Zeeman broadening grows in the same
direction, the RV signature of the Zeeman effect is significant in
infrared observations of active stars.  Thus, the RV signal of active
stars does not vanish at long wavelengths, and infrared observations
are not less affected by activity than observations at optical
wavelengths. This may be particularly important for moderately active
stars that could be populated by magnetic areas in a way similar to
the Sun \citep{2010A&A...512A..38L}.

The magnetically insensitive CO lines in the $K$-band provide a
notable exception. Their response to magnetic fields is so low that
even strong fields do not substantially distort their line
profiles. These CO lines can be very useful to disentangle RV
variations due to Keplerian orbits from magnetic
activity. Unfortunately, the CO lines are contaminated by telluric
lines introducing other problems when an accuracy on the m\,s$^{-1}$
level is desired \citep{2010ApJ...713..410B}. 

Radial velocity signatures due to convective blueshift can also be
significant in sun-like stars \citep{2010A&A...512A..39M}. Their
amplitude likely depends on line-depth and therefore adds additional
complexity to disentangling stellar activity from Keplerian signals.

It is very difficult to predict the dependence of the RV signal as a
function of wavelength because it sensitively depends on the
combination of spot temperatures and their magnetic field
strengths. Both are poorly constrained by currently available data and
simulations suggest that differences between solar and very cool star
magnetic structures exist \citep{2011ASPC..448.1071B}. Simultaneous
measurements of RV amplitude over large wavelength regions provides
useful information for characterizing stellar activity, most important
starspot temperature and magnetic fields. In moderately active stars,
the precision required for such a measurement is on the level of a few
m\,s$^{-1}$ for wavelength intervals of several hundred nm, which is a
challenge for typical RV surveys. Simultaneous RV measurements at
different wavelength bands are possible already in a few spectrographs
and will become accessible over very large ranges with high-precision
RV spectrographs operating at infrared wavelengths, as for example
CARMENES \citep{2010SPIE.7735E..37Q} and SPIRou
\citep{2011ASPC..448..771A}. Data from these facilities will not only
provide a reliable method to distinguish a Keplerian signal from
magnetic activity, they will also allow a deep look into the magnetic
and temperature structure of stellar surfaces.

\begin{acknowledgements}
  AR acknowledges research funding from DFG grant RE 1664/9-1 and support by
  the European Research Council under the FP7 Starting Grant agreement number
  279347. , DS is supported by DFG research grant RE 1664/7-1, GAE by the
  German Federal Ministry of Education and Research under 05A11MG3, and MZ by
  DFG research grant RE 1664/4-1. JM acknowledges funding as a Humboldt
  fellow. OK is supported by grants from the Knut and Alice Wallenberg
  Foundation and the Swedish Research Council.
\end{acknowledgements}

\bibliographystyle{aa}
\bibliography{refs}

\end{document}